\documentclass[amsmath,amssymb,a4paper,nofootinbib,twocolumn,pra]{revtex4}
\usepackage{graphicx}
\usepackage{bm}
\usepackage{mathbbol}
\usepackage[pdfpagemode=None,pdfstartview=FitH]{hyperref}
\newcommand{\be}{\begin{eqnarray} \begin{aligned}}
\newcommand{\ee}{\end{aligned} \end{eqnarray} }
\newcommand{\bc}{\begin{center}}
\newcommand{\ec}{\end{center}}
\newcommand{\half}{\frac{1}{2}}
\newcommand{\ran}{\rangle}
\newcommand{\lan}{\langle}
\newcommand{\im}{\text{\large $\mathbb{1}$}}
\newcommand{\Cn}{\mathbb{C}}
\newcommand{\re}{\mathop{\mathbb{R}}\nolimits}
\newcommand{\esp}{\mathop{\mathbb{E}}\nolimits}
\newcommand{\rlprt}{\mathop{\mathrm{Re}} \nolimits}
\newcommand{\imprt}{\mathop{\mathrm{Im}}\nolimits}
\newcommand{\tr}{\mathop{\mathrm{tr}}\nolimits}
\newcommand{\Tr}{\mathop{\mathrm{Tr}}\nolimits}
\newcommand{\eval}[2]{ #1  \arrowvert _{#2}}

\newcommand{\iu}{i}

\newcommand{\e}{e}

\newtheorem{examp}{Example}[section]

\newcommand{\hil}{\mathcal{H}}
\newcommand{\allrhos}{\mathcal{M}}
\newcommand{\su}{\mathfrak{su}}
\newcommand{\wavyline}[3]{
\multiput(#1,#2)(4,0){#3}{
\qbezier(0,0)(1,1)(2,0)
\qbezier(2,0)(3,-1)(4,0)}}

\begin{document}
\bibliographystyle{h-physrev}
\title{Entanglement is not very useful for estimating  multiple phases}
\author{Manuel A. Ballester}
\email{ballester@math.uu.nl}
\homepage{http://www.math.uu.nl/people/balleste/}
\affiliation{Department of Mathematics, University of Utrecht, Box
80010, 3508 TA Utrecht, The Netherlands}
\begin{abstract}
The problem of the estimation of multiple phases (or of commuting unitaries) is considered. This is a sub-model of the estimation of a 
completely unknown unitary operation where it has been shown  in recent works
that there are considerable improvements by using entangled input states and entangled measurements.
Here it is shown that when estimating commuting unitaries, there is practically no advantage in using entangled input states
or entangled measurements.
\end{abstract}
\maketitle
\section{Introduction}

\begin{figure}[ht]
\begin{center}
\setlength{\unitlength}{.035in}
\begin{picture}(30,20)(0,0)
 \thicklines
\wavyline{-32}{10}{5} \put(-9.2,10){{\vector(1,0){0}}}
\put(-30,6){{\footnotesize Input State}}
\put(-9.5,5){\framebox(9,10){$U$}}
\put(-14,17){{{\footnotesize Q. Operation}}}
\wavyline{0}{10}{7} \put(30.8,10){{\vector(1,0){0}}}
\put(7,6){{\footnotesize Output State}}
\put(26,17){{{\footnotesize Measurement}}} \put(30.5,5){\framebox(9.5,10){M}}
 \put(40.5,10){\vector(1,0){15}} \put(42,7){{\footnotesize Classical}} \put(42,4){{\footnotesize Data}}
\end{picture}
\end{center}
\caption{Quantum process tomography.}
\label{fig:fig1}
\end{figure}
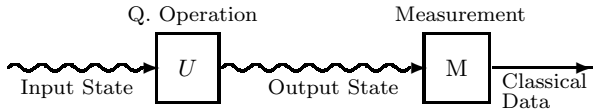

A unitary operation is a map that transforms a density operator $\rho_0$ on $\Cn^d$ to another density operator
$\rho=U\rho_0 U^{\dag}$ on  $\Cn^d$, where $U\in \text{SU}(d)$ is a $d \times d$ special unitary matrix. Suppose
one is given a device that performs an unknown $U$. One can learn something about $U$ by learning about
how it transforms a known state $\rho_0$. In order to completely determine a unitary operation one would
need to know how it transforms a basis of $\Cn^{d}$ plus some linear combinations thereof. This is known
as \emph{quantum process tomography} \cite{nielsen:book}. More precisely, to estimate $U$ one prepares 
many copies of the necessary input states and performs a measurement on the output that they produce.
As a result, some classical data are obtained and from that one can estimate $U$. This is shown schematically in Fig.\ \ref{fig:fig1}.
\medskip

\begin{figure}[ht]
\begin{center}
\setlength{\unitlength}{.035in}
\begin{picture}(30,20)(0,0)
 \thicklines
\wavyline{-32}{15}{5} \put(-9.2,15){{\vector(1,0){0}}}
\put(-30,11){{\footnotesize Entangled}}
\put(-30,8){{\footnotesize Input State}}
\put(-9.5,10){\framebox(9,10){$U$}}
\put(-14,22){{{\footnotesize Q. Operation}}}
\wavyline{0}{15}{7} \put(30.8,15){{\vector(1,0){0}}}
\put(7,11){{\footnotesize Entangled}}
\put(7,8){{\footnotesize Output State}}
\put(26,22){{{\footnotesize Measurement}}} \put(30.5,0){\framebox(9.5,20){M}}
 \put(40.5,10){\vector(1,0){15}} \put(42,16){{\footnotesize Classical}} \put(42,13){{\footnotesize Data}}
\wavyline{-32}{5}{15}\put(30.8,5){{\vector(1,0){0}}}
\end{picture}
\end{center}
\caption{Entanglement is used, $\Cn^d \otimes \Cn^d$ model.}
\label{fig:fig2}
\end{figure}
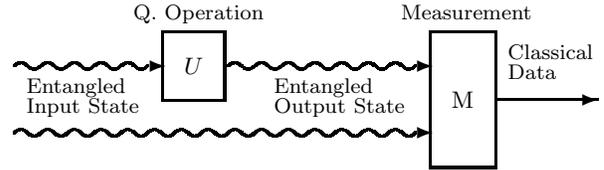

Another approach  (used in Refs.\ \cite{fujiwara:estsu2,demartini:sqd,acin:optestquantdyn,ballester:estquantop}) 
is to prepare a bipartite entangled input state 
$\rho_0$ on $\Cn^{d} \otimes \Cn^{d}$ and then use one of the parts as input for $U$, while nothing is done to the other
part, as shown in Fig.\ \ref{fig:fig2}. The effect of the operation is to transform the state $\rho_0$ to 
$(U \otimes \im) \rho_0 (U^{\dag} \otimes \im)$. 
This output state
is then measured and estimated, and, since in this case there is a one-to-one relation between the output state and $U$,
one gets an estimate for $U$ as well.
The advantages of this method with respect to quantum process tomography are that only one input state is needed, 
and that there is potentially a better accuracy in the estimation (if  nonseparable measurements are used)
\cite{ballester:estquantop}.
\medskip

In this paper a relatively less difficult problem will be studied,
the estimation of unitary operations that commute with one another, that is, only a maximal Abelian subgroup of SU$(d)$
is considered instead of the whole group.
In this case, the number of unknown parameters
decreases from $d^2-1$ to $d-1$. This problem has already been addressed in Ref.\ \cite{macchiavello:multphaseest} where it is
 given the name of ``multiple phase estimation'' (MPE).
 One would like to know whether it is also advantageous to use an entangled input in MPE. In what follows the MPE model that uses
 entanglement (Fig.\ \ref{fig:fig2}) will be referred to as MPEE and the one that does not use it (Fig.\ \ref{fig:fig1}) will be referred to as MPEU.

There are two things that need to be optimized here, the input state and the measurement that is to be performed. Therefore one
needs a quantitative measure of how good an input state is and of how good a measurement is.

\section{\label{sec:defs} Preliminaries}

In this section, the necessary concepts of  quantum Fisher information (QFI) and Fisher information
(FI) will be introduced, and the quantum Cram\'{e}r-Rao bound (QCRB) will be stated. The QFI and the 
FI will be used as measures of the performance of an input state and a measurement, respectively. The QCRB
relates these two quantities in a nice way.

\subsection{QFI}
Suppose that the quantum state density matrix $\rho$ on $\Cn^{d}$ is parametrized by $\theta \in \Theta \subset \re^p$ where $p$ is 
the number of parameters. In our case $\rho$ would be the output state and $p=d-1$. Define the symmetric logarithmic
 derivatives (SLDs) $\lambda_1,\dots,\lambda_p$ as the self-adjoint operators that satisfy
$$
\rho_{,i}(\theta)=\partial_{\theta_i}\rho(\theta)={\textstyle \half}[\rho(\theta) \lambda_i(\theta) +\lambda_i(\theta) \rho(\theta)].
$$
For pure states, $\rho=|\psi\ran \lan \psi|$, they simply are $\lambda_i=2 \rho_{,i}$.
The QFI is defined as the $p \times p$ matrix with elements
$$
H_{ij}(\theta)=\rlprt \tr \left[ \rho(\theta) \lambda_i(\theta) \lambda_j(\theta)\right]
$$
which for pure states reduces to
$$H_{ij}(\theta)=\rlprt \lan l_i(\theta) |l_j(\theta)\ran$$
where $|l_i(\theta)\ran=\lambda_i(\theta)|\psi(\theta)\ran$.

The $|l\ran$ vectors have a simple geometric interpretation. Suppose one has a pure state
model parametrized by $\theta \in \Theta \subset \re$. For simplicity, the parameter has 
been taken to be one dimensional. Denote the set of all state vectors by
$\hil=\{|\psi(\theta)\ran| \theta \in \Theta\}$ and the set of all density operators
by $\allrhos=\{\rho(\theta)|\theta \in \Theta\}$. State vectors and density operators
are related by the map $\pi:|\psi\ran \mapsto |\psi\ran \lan \psi|$, $\hil \to \allrhos$.
This map is many to one since $\pi(|\psi\ran)=\pi(\e^{i \phi}|\psi\ran)$ where
$\phi$ is an arbitrary real phase. This means that $\pi^{-1}(\rho_0)$ is a circle
in $\hil$ and that a curve in $\allrhos$ is mapped through $\pi^{-1}$ to a tube in $\hil$.
Conversely, all curves that lie on the surface of the tube, are mapped through $\pi$
to the same curve in $\allrhos$.
Then out of all curves $|\psi(\theta)\ran \in \hil$ satisfying $|\psi(0)\ran=|\psi_0\ran$ and
$\pi(|\psi(\theta)\ran)=\rho(\theta)$ ($\rho(\theta)$ is a given curve in $\allrhos$)
there is a minimal curve $|\widetilde{\psi}(\theta)\ran$, defined as the one that at every $\theta$ 
goes from the circle $\pi^{-1}[\rho(\theta)]$ to the circle $\pi^{-1}[\rho(\theta+\delta \theta)]$
using the shortest path. $|l(\theta)\ran$ is a vector pointing in the direction of that shortest path.
It can then be calculated as $|l(\theta)\ran=2 \partial_{\theta}|\widetilde{\psi}(\theta)\ran$.

\subsection{FI \label{section:fi}}
Take a positive operator values measure (POVM) with elements $M_1,\dots,M_n$. This POVM induces a probability distribution given by $p_{\xi}(\theta)=\tr \rho(\theta)M_\xi$, 
the probability to obtain outcome $\xi$ if the parameter has the value $\theta$.
The Fisher information for this measurement is defined as 
the $p \times p$ matrix with elements
$$
I_{ij}(M,\theta)=\esp_{M,\theta}[ \partial_{\theta_i}\ln p_{\xi}(\theta) \partial_{\theta_j} \ln p_{\xi}(\theta)].
$$

For an estimator $\hat{\theta}$ and a measurement $M$, locally unbiased at $\theta_0$,\footnote{This means that
the expectation of the estimator satisfies $\esp_{M,\theta_0}(\hat{\theta}_i)={\theta_0}_i$
and $\eval{\partial_{\theta_j}\esp_{M,\theta}(\hat{\theta}_i)}{\theta =\theta_0} =\delta_{ij}$.}
the (classical) Cram\'er-Rao bound is satisfied
$$V(M,\theta_0,\hat{\theta}) \geq I(M,\theta_0)^{-1},$$
i.e., the FI is the smallest variance that a locally unbiased estimator based on this measurement can have. This
also means that, if one of the eigenvalues of $I$ is zero, then the variance of the function
of the parameters corresponding to that eigenvalue is infinity and therefore cannot be
estimated.

If one has $N$ copies of the quantum state and performs the same measurement on each of the copies then the
FI of the $N$ copies, $I^N$, satisfies $I^N(M,\theta)=N I(M,\theta)$ where $I(M,\theta)$ is the FI of one system.
It follows that
$$V^N(M,\theta_0,\hat{\theta}) \geq {I^N}(M,\theta_0)^{-1}=I(M,\theta_0)^{-1}/N.$$
It is a well known fact in mathematical statistics that the maximum likelihood estimator (MLE) in the limit
of large $N$ is asymptotically unbiased and saturates the
classical Cram\'er-Rao bound. Moreover no other reasonable estimator
(unbiased or not) can do better.
\subsection{QCRB}
The QCRB states that for any measurement $M$
\be \label{eq:eqqcrb}
I(M,\theta) \leq H(\theta).
\ee
In other words, $H(\theta)-I(M,\theta)$ is a positive semidefinite matrix.

This bound is not achievable in general.
A theorem due to Matsumoto \cite{matsu:crb} states that for pure states
the bound is achievable at $\theta=\theta_0$ if and only if
\be \label{eq:condqc}
\imprt \lan l_i(\theta_0) |l_j(\theta_0)\ran=0.
\ee
If a model satisfies the above condition, it is said to be quasiclassical at $\theta_0$. Furthermore, if condition 
(\ref{eq:condqc}) holds, there is a measurement with $p+2$ elements that achieves the bound. In fact, any measurement 
of the type
\be \label{eq:optimalmeasurement}
&M_{\alpha}=|b_{\alpha}\ran \lan b_{\alpha}|,~~\alpha=1,\dots, p+1, \\
&M_{p+2}=\im-\sum_{\alpha=1}^{m+1} M_{\alpha}, \\
&|b_{\alpha}\ran=\sum_{\beta=1}^{p+1} o_{\alpha \beta}|m_{\beta}\ran,\\
&|m_k\ran=\sum_l (H^{-\frac{1}{2}})_{kl} |l_l\ran,~~
|m_{p+1}\ran=|\phi\ran,
\ee
with $o$ a $(p+1)\times(p+1)$ real orthogonal matrix satisfying $o_{\alpha,p+1}\neq 0$
achieves $I(M,\theta_0)=H(\theta_0)$.

Now one can see why these quantities are a good measure of the performance of input states and measurements. From 
$V \geq I^{-1}/N \geq H^{-1}/N$ one can see that a good input state is one that achieves an $H$ as large as possible
and a good measurement is one that achieves an $I$ as large as possible. Since $I$ and $H$  are matrices the best input state and best measurement cannot always be decided unambiguously. This ambiguity will not be completely resolved
in the case of choosing an input state. However, in the case of choosing an optimal measurement there will be no ambiguity. 
In the next section, it will be shown that  for every input state
there is a measurement on the output state that achieves equality in Eq.\ (\ref{eq:eqqcrb}). Therefore, an optimal
measurement is one that achieves equality in the QCRB.

\section{MPE is a quasiclassical model \label{section:mpeqc}}

It will be shown here that both MPEE and MPEU are quasiclassical everywhere (for all $\theta \in \re^{d-1}$) and for any input state. 
In particular, this means that only one input state is necessary in MPEU also. In this respect MPE is quite different from
 the estimation of a completely arbitrary $U$. Actually, since MPEU can be considered as a special case 
of MPEE, one needs to show quasiclassicality only for MPEE.

The MPEE model is 
\be \label{eq:model}
\rho(\theta)=U_\theta \rho_0 U^{\dag}_\theta
\ee
where
$$U_\theta=\exp\left(\iu \sum_{m=1}^{d-1}\theta_m T_m\right)\otimes \im,$$
$\rho_0$ on  $\Cn^d \otimes \Cn^d$ is a pure state density matrix,\footnote{$\rho_0$ can be taken to be pure
since it was shown in \cite{fujiwara:chanident} that the QFI is convex.} and $T_1,\dots,T_{d-1}$ are selfadjoint
traceless matrices that commute with one another. They are chosen so that they satisfy an orthonormality condition
\be\label{eq:orthnorm}
\tr T_m T_n=\delta_{nm}.
\ee

The SLDs are 
$$\lambda_m = 2 \partial_m\rho(\theta)=2 \iu [T_m \otimes \im,\rho(\theta)] $$
and 
$$
\tr\rho\lambda_m\lambda_n=4\{\tr\rho_0(T_mT_n\otimes\im)-\tr[\rho_0(T_m\otimes\im)\rho_0(T_n\otimes\im)]\}.
$$
It is easy to see that due to the commutativity of the $T$'s this quantity is real and therefore
the model is quasiclassical, i.e., it satisfies the condition (\ref{eq:condqc}). Therefore for 
every $\theta$ there exists a measurement $M$ (which may depend on $\theta$) such that $I(M,\theta)=H(\theta)$.
Furthermore, if one has a large number $N$ of copies, performs the same (optimal) measurement on all $N$, and calculates the MLE, the mean square error should behave as 
\be\label{eq:asympvar}
V^N(M,\theta,\hat{\theta}_{MLE}) =H(\theta)^{-1}/N + o(1/N).
\ee

\section{Optimal input state \label{sect:maxtrace}}
It is now clear from Eq.\ (\ref{eq:asympvar}) that one needs to choose the input state so that the QFI is ``large.''
Suppose there is an input state that has a QFI that is larger than or equal to the QFI of any other 
input state. In that case one can unambiguously choose that state as the optimal one. Unfortunately,
in our case there is not such a state. Furthermore, there are situations in which one has two input states
$\rho_1$ and $\rho_2$ with QFI $H_1$ and $H_2$, respectively, which satisfy neither $H_1 \leq H_2$ nor $H_1 \geq H_2$.
This is resolved by maximizing a quantity like $\Tr G H$, with $G$ a real positive semidefinite matrix. By doing 
this one assigns relative weights to the mean square errors of the different parameters. Furthermore, 
achieving maximum $\Tr G H$ ensures that no other input state can  have a larger QFI.
In this paper a particular choice
is made: all parameters are given the same importance, i.e., the input state will be chosen so that it maximizes $\Tr H$. 
With this particular choice it is possible to obtain nice analytic results and, since the trace of the QFI is parametrization invariant,
the results obtained will not depend on the chosen parametrization.
\subsection{MPEE}
Since the self-adjoint matrices $T_1,\dots,T_{d-1}$ commute with one another, there is a basis where all of them are diagonal $\{|1\ran,  \dots, |d\ran\}$ (this basis is considered to be known).
From this point, all calculations will be made in this basis.

The input state is $\rho_0=|\Psi_0 \ran \lan \Psi_0|$, and $|\Psi_0 \ran$ can be expanded as $|\Psi_0\ran = \sum_{kl} R_{kl}|kl\ran$. The partial trace is then $\tr_B |\Psi_0\ran \lan \Psi_0|=RR^{\dag}$, and since $RR^{\dag}$ is self-adjoint and has trace $1$, it can be written as
$$RR^{\dag}=\frac{\im}{d}+\sum_{\alpha=1}^{d^2-1}t_\alpha T_{\alpha},$$ 
where, in general,  the last sum includes  all generators of the Lie algebra $\su(d)$. Then the QFI is

\be
H_{mn}&=4\left[ \lan \Psi_0|(T_m T_n \otimes \im)|\Psi_0\ran \right.  \\
&-\left. \lan \Psi_0|(T_m \otimes \im)|\Psi_0\ran \lan \Psi_0|( T_n \otimes \im)|\Psi_0\ran \right] \\
&=4\left[\tr (RR^{\dag}T_mT_n)-(\tr RR^{\dag}T_m)(\tr RR^{\dag}T_n)\right] \\
&=4\left[\tr (RR^{\dag}T_m T_n)-t_m t_n\right]
\ee
and its trace is
$$\Tr H = 4\left[\tr\left(RR^{\dag}\sum_{m=1}^{d-1}T_m^2\right)-\sum_{m=1}^{d-1}t_m^2\right].$$
The commuting $T$'s can be written as $T_m=\sum_{k=1}^{d}c_{mk}|k\ran \lan k|$. The tracelessness condition implies 
$\sum_{k=1}^d c_{mk}=0$,
while the orthonormality condition (\ref{eq:orthnorm}) implies
$\sum_{k=1}^{d}c_{mk} c_{nk}=\delta_{mn}$.
These two together lead to $$\sum_{m=1}^{d-1}c_{mk}c_{ml}=\delta_{kl}-\frac{1}{d}$$
and then to
$$
\sum_{m=1}^{d-1}T_m^2=\sum_{k=1}^d \sum_{m=1}^{d-1}c_{mk}^2 |k \ran \lan k|=\frac{d-1}{d}\im.
$$
Substituting this in the equation for the trace, one gets
$$\Tr H = 4\left[\frac{d-1}{d}-\sum_{m=1}^{d-1}t_m^2\right],$$
and one immediately sees that $\tr H$ is maximal if and only if $t_m=0$, $m=1,\dots,d-1$. 
Of course the rest of the $t$'s
can be anything (as long as $RR^{\dag}$ remains positive).

Concluding, any pure input state $\rho_0$ satisfying
$$
\tr \rho_0(T_m \otimes \im)=0,~\forall~m=1,\dots,d-1,
$$
achieves a QFI $H_{mn}=4~\delta_{mn}/d$ which has the maximum possible trace
among all QFIs. In particular, maximally entangled states satisfy this condition.

In the full SU$(d)$ model one obtains a similar result, and the maximum is
attained at and only at the maximally entangled state; 
this will be shown in appendix \ref{appendix:appa}.
\subsection{MPEU}
It will be shown here that for every entangled input state 
$|\Psi_0\ran=\sum_{kl}R_{kl}|kl\ran \in \Cn^d \otimes \Cn^d$ and every 
value of the ($d-1$)-dimensional parameter $\theta$ there is an input state $|\psi_0\ran \in \Cn^d$ that achieves 
the same QFI everywhere. This means that in this model, unlike in the full SU($d$) model, there is no improvement in the accuracy of the estimation
by using entangled inputs.

The model is now 
$$\rho(\theta)=U_\theta \rho_0 U^{\dag}_\theta$$
where
$$U_\theta=\exp\left(\iu \sum_{m=1}^{d-1}\theta_m T_m\right)$$
and $\rho_0 = |\psi_0 \ran \lan \psi_0|$ is a pure state density matrix.
Of course this model is also quasiclassical and 
by a calculation identical to the one made in the $\Cn^d \otimes \Cn^d$
case one gets that the QFI is
$$H_{mn}=4\left[ \lan \psi_0|T_m T_n|\psi_0\ran -\lan \psi_0|T_m|\psi_0\ran \lan \psi_0| T_n|\psi_0\ran \right].$$
It is not difficult to see that any input state of the form
$$
|\psi_0\ran=\sum_{k=1}^d \sqrt{\lan k|RR^{\dag}|k\ran}~ \e^{\iu \phi_k}|k\ran
$$
where the $\phi$'s are arbitrary phases, achieves 
\be \nonumber
H_{mn}&=4\left[ \lan \Psi_0|(T_m T_n \otimes \im)|\Psi_0\ran \right.\\
&-\left. \lan \Psi_0|(T_m \otimes \im)|\Psi_0\ran \lan \Psi_0|( T_n \otimes \im)|\Psi_0\ran \right],
\ee
the QFI when the input is the entangled state $|\Psi_0\ran$. 

In particular, any input state of the form
\be \label{eq:optimalinputstate}
|\psi_0\ran=\frac{1}{\sqrt{d}}\sum_{k=1}^d  \e^{\iu \phi_k}|k\ran
\ee
achieves the maximum trace of the QFI. A state of this form (with all the
$\phi$'s set to zero) was used in Ref.\ \cite{macchiavello:multphaseest}.

\section{Optimal measurement}
\subsection{MPEU}
This model (and actually also the MPEE) has the property that 
$I(\theta,U_{\theta}MU_{\theta}^{\dag})=I(0,M)$, which is easy to prove; the only necessary ingredient is that 
$U_{\theta}^{\dag}\partial_{\theta_k} U_{\theta}=\partial_{\theta_k}U_{\theta}|_{\theta=0}$. Therefore,
if one has an optimal measurement $M$ at $\theta=0$,  then the measurement $U_{\theta}MU_{\theta}^{\dag}$ will be optimal at $\theta$.

One can now use the recipe given by Eq.\ (\ref{eq:optimalmeasurement}) to find an optimal measurement at $\theta=0$.
The optimal input state\footnote{i.e. the one that achieves maximum trace of the QFI} given by Eq.\ (\ref{eq:optimalinputstate}) is used. At the origin one has
\be \nonumber
|l_n\ran &= \frac{2 \iu}{\sqrt{d}}\sum_{k=1}^d c_{nk}~\e^{i \phi_k} |k\ran,\\
|\psi\ran &= \frac{1}{\sqrt{d}}\sum_{k=1}^d \e^{i \phi_k} |k\ran.
\ee
From these vectors one can form an orthonormal set
\be \nonumber
|m_n\ran &= \iu\sum_{k=1}^d c_{nk}~\e^{i \phi_k} |k\ran, ~n=1,\dots,d-1,\\
|m_d\ran &= \frac{1}{\sqrt{d}}\sum_{k=1}^d \e^{i \phi_k} |k\ran
\ee
and with the choice $o_{kl}=\delta_{kl}-2/d$ one gets the set of orthonormalized states onto which 
the measurement elements will project,
\be
|b_k\ran =|m_k\ran -\frac{2}{d}\sum_{l=1}^{d}|m_l\ran,~~k=1,\dots,d.
\ee
The optimal measurement at $\theta=0$ has elements $M_k=|b_k\ran \lan b_k|$ and since the above vectors form an orthonormal 
basis of $\Cn^d$, they satisfy $\sum_{k=1}^d M_k=\im$. The optimal measurement at $\theta$ has elements 
$U_{\theta}|b_k\ran \lan b_k| U_{\theta}^{\dag}$.

Note that the above choice of the orthogonal
matrix $o$ works for $d\geq 3$; for $d=2$ another matrix must be chosen. The $d=2$ case will be
treated in the following example.
\begin{examp}[$d=2$]
The orthonormal set formed from the input state and $|l\ran$ is 
\be
|m_1\ran=\frac{\iu}{\sqrt{2}}(|0\ran-|1\ran)\\
|m_2\ran=\frac{1}{\sqrt{2}}(|0\ran+|1\ran).
\ee
These vectors are then rotated to obtain
\be
|b_1\ran = \cos{\eta} |m_1\ran - \sin{\eta} |m_2\ran 
=\frac{\iu}{\sqrt{2}}(\e^{\iu \eta}|0\ran-\e^{-\iu \eta}|1\ran) \\
|b_2\ran = \sin{\eta} |m_1\ran + \cos{\eta} |m_2\ran 
=\frac{1}{\sqrt{2}}(\e^{\iu \eta}|0\ran+\e^{-\iu \eta}|1\ran),
\ee
where $\eta$ must satisfy $\sin{\eta}\neq 0$ and $\cos{\eta}\neq 0$. The
measurement elements are $|b_1\ran \lan b_1|$ and $|b_2\ran \lan b_2|$.
The FI of this measurement at the origin is equal to the QFI (equal to $2$) 
as expected. Furthermore, this equality happens to hold everywhere and not only
at the origin. This feature is very useful in practice, it means that the 
optimal measurement does not depend on the actual (unknown) value of the
parameter and therefore an adaptive scheme is not necessary. Whether 
it is possible to find measurements with this characteristic for 
$d>2$ is an open problem.
\end{examp}
\subsection{MPEE}
An optimal measurement in this case can also be derived using the recipe given by Eq.\ (\ref{eq:optimalmeasurement}).
In general, such a measurement is a joint measurement on the two output systems. One could ask whether it is possible 
to achieve the bound with local measurements and classical communication. In what follows, it will be shown that 
this is indeed possible.

The input state is taken to be the maximally entangled state $\sum_{k=1}^d|kk\ran/\sqrt{d}$. Alice measures the system coming out
of $U$ and Bob measures the other one. The strategy is the following. Bob performs the von Neumann measurement
$B_k=|w_k\ran \lan w_k|$ with $\sum_k B_k=\im$ on his system where 
$$
|w_k\ran = \frac{1}{\sqrt{d}}\sum_{l=1}^d \exp\left(\frac{2 \pi kl}{d}\iu\right)~~|l\ran,~k=1,\dots,d.
$$
He obtains outcome $k$ with probability $1/d$; he then phones Alice and tells her the outcome of his measurement. The net result 
of this is that Bob prepares the state 
$$
 \frac{1}{\sqrt{d}}\sum_{l=1}^d \exp\left(-\frac{2 \pi kl}{d}\iu\right)~|l\ran
$$
at the input of $U$. It is easy to recognize this state as one of the optimal states in the $\Cn^d$ case. Now Alice
can perform the measurement $A_{kl}$ with $\sum_l A_{kl}=\im$, which is the optimal measurement described above for the $\Cn^d$
case, and where the arbitrary phases are now fixed to $\phi_l=2 \pi k/d$. It is not difficult to check that this measurement
indeed achieves equality in the QCRB at $\theta=0$. The measurement on $\Cn^d \otimes \Cn^d$ is then 
$\sum_{kl}A_{kl}\otimes B_k=\im \otimes \im$ and the measurement
$\sum_{kl} U_{\theta} A_{kl} U_{\theta}^{\dag}\otimes B_k=\im \otimes \im$ is optimal at $\theta$.

If this is applied to the $d=2$ case and one uses the $\theta$-independent 
optimal measurement derived for MPEU, a $\theta$-independent optimal 
measurement is obtained for MPEE.
\section{Asymptotic Fidelity}

From the results obtained previously, one can infer the asymptotic behavior of
the average fidelity. Indeed, in Ref.\ \cite{hubner:excompbures}, it was
established that the fidelity between nearby states is given by\footnote{Actually, this fidelity is the 
square of the fidelity used in Ref.\ \cite{hubner:excompbures}.}
\be
\mathcal{F}(\theta,\theta+\delta\theta)=&
\left(\tr \sqrt{\sqrt{\rho(\theta)}\rho(\theta+\delta \theta)\sqrt{\rho(\theta)}}\right)^2\\
= &1 - \sum_{\alpha,\beta=1}^p \frac{H(\theta)_{\alpha \beta}}{4}\delta \theta_\alpha \delta
\theta_\beta +o(\delta \theta^2),
\ee
where $p$ is the number of parameters and $H$ is the QFI. In the case studied here, this fidelity 
would be between output states.

Denote by $\hat{\theta}_{\xi}$ the guess for $\theta$ if the outcome of the measurement
was $\xi$, then the fidelity, averaged over all possible outcomes, is

\be
F(\theta,\hat{\theta})&=\sum_{\xi} \tr[\rho(\theta)M_\xi]~\mathcal{F}(\theta,\hat{\theta}_\xi)\\
&= 1 - \frac{\tr H(\theta)V(M,\theta,\hat{\theta})}{4}+o(\delta \theta^2),
\ee
where $V(M,\theta,\hat{\theta})_{\alpha \beta}=
\sum_\xi \tr [\rho(\theta)M_\xi]~(\hat{\theta}_{\xi\alpha}-\theta_\alpha)
(\hat{\theta}_{\xi\beta}-\theta_\beta)$ is the mean square error of measurement $M$ and
estimator $\hat{\theta}$.

\begin{figure}[ht]
\begin{center}
\includegraphics[height=5cm]{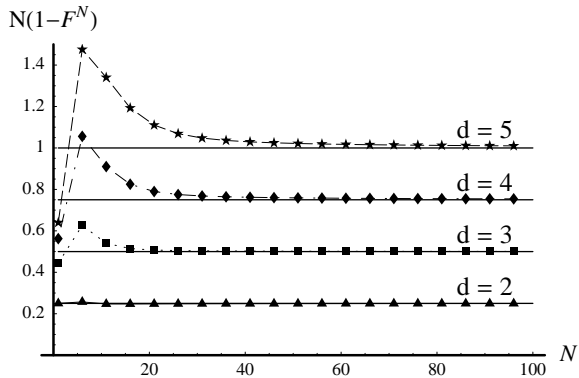}
\end{center}
\caption{The points are $N(1-F^N)$ as a function of $N$ for $d=2,\dots,5$, where
$F^N$ is the optimal fidelity obtained in  Ref.\ \cite{macchiavello:multphaseest}. The
continuous lines are at the value $(d-1)/4$ for $d=2,\dots,5$, ($0.25, 0.5, 0.75$, and
$1$, respectively.)}
\label{fig:fig3}
\end{figure}

Therefore, using Eq.\ (\ref{eq:asympvar}) one gets
\be \label{eq:asympfid}
\lim_{N \to \infty} N[1-F(\theta,\hat{\theta})^N)]=\frac{\Tr H(\theta)H(\theta)^{-1}}{4}
=\frac{d-1}{4}.
\ee
This result can be compared with the optimal fidelities obtained in Ref.\
\cite{macchiavello:multphaseest}. These optimal fidelites were also averaged
with respect to a uniform prior distribution on $\theta$. However since the result obtained 
here does not depend on $\theta$, its average with respect to any prior will be 
itself. The comparison is shown	 in Fig.\ \ref{fig:fig3}.

One can easily see that for large $N$ the optimal fidelity of  Ref.\ \cite{macchiavello:multphaseest}
 agrees with the result obtained here. Actually, this can also be proved analytically but the proof 
 will not be shown here.
\section{Conclusions}
Two models have been compared, the model of estimating commuting unitaries with and without the use of
entangled inputs (MPEE and MPEU, respectively).

It has been shown that the quantum Cr\'amer-Rao bound is achievable in both MPEE and MPEU. It has also 
been shown that any quantum Fisher information matrix  that can be attained in MPEE can also be
achieved in MPEU.
These two facts imply that an entangled input state is unnecessary.
A condition for attaining maximal trace of the QFI has been derived.

In the MPEE it has also been shown that there is a separable measurement
that achieves equality in the QCRB. 

In the $d=2$ case, measurements that are optimal everywhere have been found in both MPEU and MPEE.
This is a useful feature in practice since this means that an adaptive scheme
would not be necessary in this case. However, it is unclear whether it
is possible to find measurements with this characteristic in
general. It is also an open question whether entanglement could prove
itself useful in this respect (for $d>2$).

These facts show that entanglement is, at best, not as useful in estimating commuting unitaries
as in the estimation of a completely unknown unitary.
\begin{acknowledgments}
This research was funded by the Netherlands Organization for
Scientific Research (NWO), support from the RESQ (Grant No.\ IST-2001-37559)
project of the IST-FET programme of the European Union is also
acknowledged.
\end{acknowledgments}

\appendix
\section{Generalization of the result in section \ref{sect:maxtrace} to the full model \label{appendix:appa}}

In this appendix, a result similar to that of Sec.\ \ref{sect:maxtrace} will be proved
in the model that includes the whole of SU($d$) and not only a commuting subgroup, i.e., the model
considered in Ref.\ \cite{ballester:estquantop}.

Denote by $H^{\rho_0}(\theta)$ the QFI at $\theta$ if 
the input state is $\rho_0$, and by $\widetilde {H}$ the QFI when the input state is a maximally entangled state; in what follows the dependence on $\theta$ will be omitted. Then the following inequality holds for any input state $\rho_0$:
\be
\Tr (\widetilde{H}^{-1}H^{\rho_0}) \leq d^2-1,
\ee
and equality is attained if and only if $\rho_0$ is a maximally entangled state. This will be proved in what follows.

Notice that this trace is parametrization invariant, and not just $\Tr H^{\rho}$ as in Sect.\ \ref{sect:maxtrace}, in that case  $\widetilde H$ was proportional to the identity so there is no contradiction.

The model is again described by Eq.\ (\ref{eq:model}) but now $U$ is 
$$U_\theta=V_{\theta} \otimes \im$$
where $V_{\theta}=\exp\left(\iu \sum_{\alpha=1}^{d^2-1}\theta_{\alpha} T_{\alpha}\right)$. 
As before, the $T$'s are traceless self-adjoint matrices chosen so that $\tr (T_\alpha T_\beta)=\delta_{\alpha \beta}$.
The input state $\rho_0$ is chosen to be pure because of the convexity of the QFI \cite{fujiwara:chanident}.
The SLDs are 
$\lambda_\alpha(\theta)=2 \rho{,_\alpha}(\theta)$, where $ \rho{,_\alpha}(\theta)$ means the partial derivative of $\rho(\theta)$
with respect to $\theta_\alpha$. The matrix elements of $H^{\rho_0}$ are
$$
H^{\rho_0}_{\alpha \beta}=\rlprt \tr[ \rho \lambda_\alpha \lambda_\beta] = 
4 \rlprt \tr[ \rho \rho_{,\alpha} \rho_{,\beta}].
$$
Since $\rho_0$ is pure it can be written as  $\rho_0=|\psi_0 \ran \lan \psi_0|$ and $|\psi_0 \ran=\sum_{kl}R_{kl}|kl\ran$.
$H^{\rho_0}_{\alpha \beta}(\theta)$ can then be calculated to be
\be \nonumber
H^{\rho_0}_{\alpha \beta}&=4 \rlprt \left [\tr(RR^{\dag}V_{,\alpha}^{\dag}V_{,\beta})\right. \\
&+\left. \tr(RR^{\dag}V^{\dag}V_{,\alpha})\tr(RR^{\dag}V^{\dag}V_{,\beta}) \right].
\ee
Denote $S_{\alpha}=-\iu V^{\dag}V_{,\alpha}$; then 
$$
H^{\rho_0}_{\alpha \beta}=4 \rlprt \left [\tr(RR^{\dag}S_{\alpha}S_{\beta}) -
\tr(RR^{\dag}S_{\alpha})\tr(RR^{\dag}S_{\beta}) \right].
$$
Note that $S_\alpha\in\mathfrak{su}(d)$. 
Substituting $RR^{\dag}=\im/d$ in the expression for $H^{\rho_0}$; one gets 
$$
\widetilde{H}_{\alpha \beta}=\frac{4}{d}\tr(S_{\alpha}S_{\beta}).
$$
The matrices $S_1,\dots,S_{d^2-1}$ can be orthonormalized,
$$
\tr\left[\left( \frac{2}{\sqrt{d}}\sum_\mu \widetilde{H}^{-1/2}_{\alpha \mu}S_{\mu} \right)  
\left(\frac{2}{\sqrt{d}}\sum_\nu \widetilde{H}^{-1/2}_{\beta \nu}S_{\nu}\right)
 \right]=\delta_{\alpha \beta}.
$$
The operator
\be \nonumber
\sum_\alpha \left(\frac{2}{\sqrt{d}}\sum_\mu \widetilde{H}^{-1/2}_{\alpha \mu}S_{\mu}\right)  
\left(\frac{2}{\sqrt{d}}\sum_\nu \widetilde{H}^{-1/2}_{\alpha \nu}S_{\nu}\right)\\
=\frac{4}{d}\sum_{\mu \nu} \widetilde{H}^{-1}_{\mu \nu}S_{\mu}S_{\nu}
\ee
is a Casimir operator and therefore proportional to the identity. The proportionality factor 
can be found by taking the trace, and finally one gets
$$
\sum_{\mu \nu} \widetilde{H}^{-1}_{\mu \nu}S_{\mu}S_{\nu}=\frac{d^2-1}{4}\im.
$$

The wanted trace is
$$
\Tr(\widetilde{H}^{-1}H^{\rho_0})=d^2-1 - \sum_{\alpha \beta} \widetilde{H}^{-1}_{\alpha \beta} 
\tr(RR^{\dag}S_{\alpha})\tr(RR^{\dag}S_{\beta}).
$$
This quantity is always less than or equal to $d^2-1$ and, furthermore, this value is attained
if and only if $\tr(RR^{\dag}S_{\alpha})=0$ for all $\alpha=1,\dots,d^2-1$, which implies that
$RR^{\dag}=\im / d$, i.e., $\rho_0$ is maximally entangled. In particular, this implies
that there is no input state $\rho_0$ for which $H^{\rho_0} \geq \widetilde{H}$.
\begin{flushleft}

\end{flushleft}
\end{document}